# Understanding the Prediction Mechanism of Sentiments by XAI Visualization

Chaehan So
Information & Interaction Design
Humanities, Arts & Social Sciences
Division, Yonsei University
Seoul, South Korea
Email: cso@yonsei.ac.kr

## ABSTRACT

People often rely on online reviews to make purchase decisions. The present work aimed to gain an understanding of a machine learning model's prediction mechanism by visualizing the effect of sentiments extracted from online hotel reviews with *explainable AI (XAI)* methodology.

Study 1 used the extracted sentiments as features to predict the review ratings by five machine learning algorithms *(knn, CART decision trees, support vector machines, random forests, gradient boosting machines)* and identified random forests as best algorithm.

Study 2 analyzed the random forests model by feature importance and revealed the sentiments joy, disgust, positive and negative as the most predictive features. Furthermore, the visualization of additive variable attributions and their prediction distribution showed correct prediction in direction and effect size for the 5-star rating but partially wrong direction and insufficient effect size for the 1-star rating. These prediction details were corroborated by a what-if analysis for the four top features.

In conclusion, the prediction mechanism of a machine learning model can be uncovered by visualization of particular observations. Comparing instances of contrasting ground truth values can draw a differential picture of the prediction mechanism and inform decisions for model improvement.

## Keywords
Explainable AI; interpretable AI; sentiment analysis

## 1. INTRODUCTION

Sentiment analysis has established itself as a common natural language processing (NLP) approach to analyze online media, e.g. for the prediction of stock prices [1] or overall market trend [2].

To predict commercial success, research has further retrieved sentiments from movie reviews [3], restaurant reviews [4], or hotel reviews [5]. Retrieving sentiments from online reviews is relevant because purchase decisions rely on people's recommendations given in such online reviews.

Most sentiment analysis models are based on the theory of *basic emotions* by Ekman [6] that postulates the existence of a set of universal (cross-cultural and inherited) emotions including happiness, surprise, fear, sadness, anger, and disgust. Neuroscientific research has validated these basic emotions by corresponding neural correlates [7].

Apart from basic emotion, *emotional valence*, the positive or negative direction of an emotion, is a common category in sentiment analysis. Psychological research has emphasized that positive and negative valence are not mutually exclusive but can be experienced simultaneously and at different levels – in other words, positive and negative emotions are independent [8].

Although machine learning research has extensively applied sentiment analysis, it is surprising that until thus far, it has not tried to understand the underlying prediction mechanism of the models. Such investigation was long prevented by the notion of machine learning algorithms being a "blackbox". This notion was closely related to the wrong assumption that a machine learning model's prediction mechanism cannot be uncovered. In this regard, a new research field has emerged around the terms *interpretable AI* and *explainable AI (XAI)* [9], [10] with a set of new methods. These XAI methods aim to disentangle the details of what a machine learning model has learned during model training, and how it applies that knowledge to the prediction of particular observations. Visualizing these details can create insights that are difficult to reach by an analytical approach, especially when the effect of several features shall be compared for different observations. The visualization of such differential effects can bring an intuitive understanding of the underlying relationships even in the presence of many data points.

Taken together, the preceding considerations lead to a research question that can be formulated as follows:

> How can correct and incorrect predictions of a machine learning model be visualized to explain its prediction mechanism?

In the quest for an answer to this research question, the current work extracted sentiments from online hotel reviews, and used them as features to train five different machine learning models. Then, it analyzed the best performing model by various XAI methods to identify the correct and incorrect prediction cases, and thereby to gain an understanding of the machine learning model's prediction mechanism.

## 2. Method

The analysis was performed on a dataset of 10.000 consumer reviews on hotels in the US gathered by the data company Datafiniti between December 2018 and May 2019.

### 2.1.1 Data Preprocessing

The data preprocessing encompassed the steps

a) removing variables with uniform distribution (country had 100% value "US", primaryCategories had 99.9% value "Accomodation & Food Services")
b) removing variables with more than 10% missing values (18.1% in postalCode)
c) removing technical information (id, dateAdded, dateUpdated, keys, reviews.date, reviews.dateSeen, reviews.sourceURLs, reviews.username, sourceURLs, websites)

### 2.1.2 Descriptive Statistics

The variable *primaryCategories* contained 99.9% *Accomodation & Food Services*. The top 10 cities and provinces for the hotels are displayed in Table 1.

**Table 1. Top 10 Hotel Cities and Hotel Provinces**

| Rank | Hotel City Name | Percentage | Hotel Province | Percentage |
|---|---|---|---|---|
| 1 | San Diego | 11.90% | California | 26.50% |
| 2 | San Francisco | 8.08% | Florida | 12.8% |
| 3 | New Orleans | 7.98% | Georgia | 8.44% |
| 4 | Atlanta | 7.62% | Louisiana | 8.17% |
| 5 | Orlando | 7.34% | Washington | 7.30% |
| 6 | Seattle | 6.35% | Texas | 5.59% |
| 7 | Chicago | 4.59% | Illinois | 4.87% |
| 8 | Honolulu | 3.05% | Hawaii | 3.65% |
| 9 | Dallas | 2.45% | Pennsylvania | 2.81% |
| 10 | Anaheim | 2.36% | New York | 1.58% |

The five most frequented hotels in this dataset were the Hyatt House Seattle/Downtown (2.09%), Hotel Emma (1.83%), French Market Inn (1.44%), St. James Hotel (1.36%), and the Grand Hyatt Seattle (1.35%).

The users who wrote the hotel reviews mainly came from California (9.02%), Florida (5.34%), Texas (5.33%), Canada (3.7%), United Kingdom (3.65%), and New York (3.07%). Their main city origins were New York City (1.71%), Los Angeles (1.57%), Chicago (1.37%), Houston (1.36%), and San Diego (1.31%). No further demographic information about the users (e.g. age, gender, ethnicity, or income) was available in this dataset.

### 2.1.3 Sentiment Analysis

The sentiments were extracted by mapping the preprocessed review ratings text with the NRC sentiment lexicon [11], [12]. With 13901 word-to-sentiments mappings, it represents one of the largest lexicon corpora available. Furthermore, it provides emotion valence (negative, positive) in addition to eight basic emotions (anger, anticipation, disgust, fear, joy, sadness, surprise, trust). The sentiment extraction, and creating a hold-out set (n = 100) for the instance-level analysis, led to a reduced sample size of 9825.

### 2.2 Study Design

The dataset was analyzed in three subsequent studies. Before these studies, a prescreening of the features was performed.

Study 1 conducted model training as regression and classification (with the target variable converted into five class levels).

Study 2 applied XAI methods to analyze the best model's prediction on a global and local level.

### 2.3 Benchmarking Method

The present work compared the R implementations of the machine learning algorithms knn for k-nearest neighbors [13], svmRadial (method ksvm from package kernlab with radial kernel) [14] for support vector machines, rpart for CART decision trees [15], rf (package randomForest) [16] for Random Forests, and gbm [17] for Gradient Boosting Machines.

### 2.4 Explainable AI (XAI)

As machine learning algorithms have become ubiquitous in data science, a general skepticism emerged about their reliability and whether their blackbox nature was indeed impenetrable. This notion lead to the demand for *interpretable AI*, i.e. the "ability to explain or to present in understandable terms to a human" [18]. New methods of interpretable AI were developed in recent years [18]. Synonymously, the term *explainable AI (XAI)* has been used increasingly.

This work applies several XAI methods as provided by the DALEX package [19]. To understand these methods, it is important to note that most of them work on the *instance level*, i.e. on one or few observations, as opposed to the *global level*, i.e. all observations (e.g. the entire training set). The selection of methods presented in this work is neither exhaustive nor representative for XAI methodology. Nevertheless, this selection serves the purpose of gaining a better understanding for the functioning of the prediction mechanism learned by a machine learning algorithm.

### 2.4.1 Feature Importance

Across all observations, the *feature importance* shows a model's differential use of the features to calculate its prediction. The feature importance is a numeric value of a feature' prediction impact, i.e. its global relevance for generating the prediction in the trained model. This relevance is model-dependent, i.e. for tree-based models like gradient boosting machines or random forests, it corresponds to its role in splitting the trees, whereas for linear models, it corresponds to the normalized regression coefficient. Apart from that, feature importance can also be calculated by a model-agnostic approach using the loss function difference between a validation set and permutations of one feature in this validation set [20].

### 2.4.2 Additive Variable Attributions

The feature importance is calculated across all observations, and does not account for any variability. Nevertheless, the feature influence on specific instances of the dataset may vary distinctly from the average behavior. Such an analysis is particularly useful for comparing predictions for distinctly different ground truth values.

The contribution of a particular feature to the prediction of a machine learning algorithm is additive for tree-based models like random forests or gradient boosting machines. An instance-level analysis of additive variable attributions can be performed by various methods described as follows.

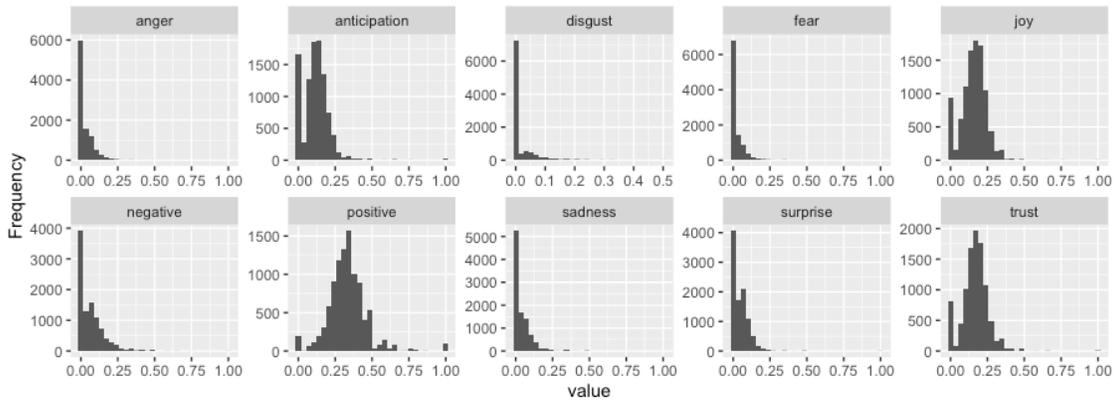

**Figure 1: Histogram of sentiment features**

A *breakdown plot* visualizes the additive nature by showing the stepwise contribution of each feature to the prediction of a single observation (instance). Each feature's contribution or variable attribution can be of positive or negative direction, and of different effect size. The sum of all additive attributions is equal to the score of the model's prediction for the particular instance.

Alternatively, the breakdown plot can show the mean prediction and distribution of all predictions (by violin plots) for each feature with the value of a particular observation. The change in prediction with each step of the next feature, with the instance's feature value assigned to, is shown by connected lines.

### 2.4.3 What-if Analysis

The ceteris-paribus plot shows the change of the target variable in response to a change of a particular feature under *ceteris-paribus* i.e. all other conditions (here: feature values) held constant.

The ceteris-paribus condition allows a so-called *what-if analysis* – seeing the effect of a single feature. The ceteris-paribus plot can thus be used to compare its feature-response relationships with the contribution direction and effect size in the breakdown plots.

The most important features (by effect size) from the breakdown and ceteris-paribus plots can be compared with their ranking in the global feature importance. This allows the conclusion about how consistent the instance-based prediction is with the overall prediction mechanism.

## 3. Results

### 3.1 Prescreening

Before the main studies, the current work performed a pre-screening of the preprocessed dataset.

The histograms in Figure 1 show that the data distribution was very different across features. The sentiments joy, positive and trust were approximately normal distributed. In contrast, the sentiments anger, disgust, fear, negative, sadness, and surprise showed a distinct floor effect.

The correlation plot in Figure 2 reveals the presence of linear relationships with the reviews rating. Except the sentiments anticipation and surprise, all remaining sentiments show moderate linear relationships ranging between 0.3 and 0.4 on average.

### 3.2 Study 1

The benchmarking results for the regression task are displayed in Figure 3. Random forests achieve the lowest RMSE (mean = 0.929, sd = 0.0175), closely followed by gbm (mean = 0.932, sd = 0.0168). The remaining algorithms show distinctly lower performance.

Comparing the regression results with a second benchmarking, performed as classification task, shows similar results. Random forests yield the best prediction with a mean accuracy of 51.9%, but are closely followed by gbm (51.5%) support vector machines (51.5%), and rpart (51.2%).

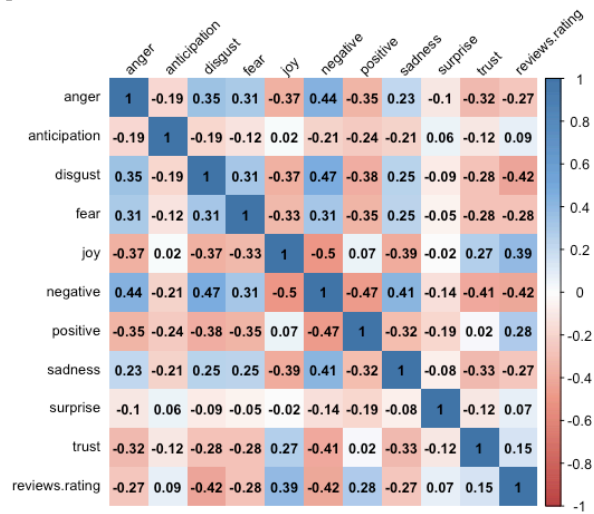

**Figure 2: Correlation Plot**

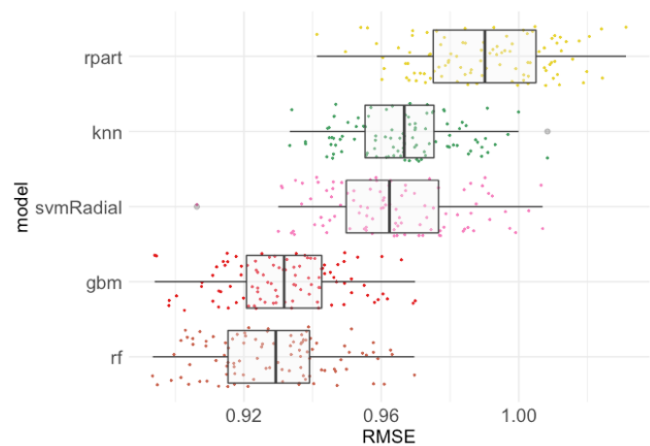

**Figure 3: Machine Learning Models Benchmarking**

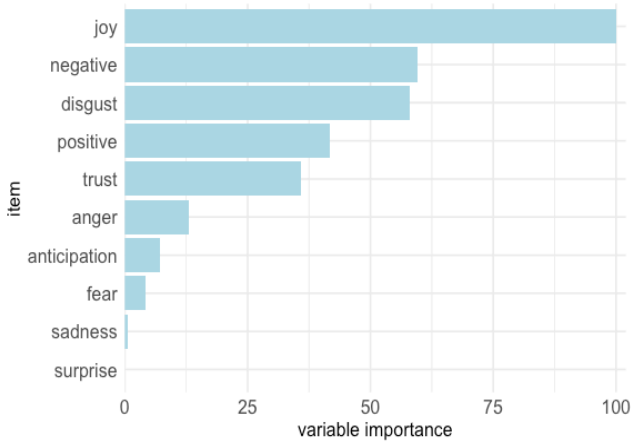

Figure 4: Global feature importance

As expected, knn yield the worst prediction with an accuracy (48.5%) reaching only the level of the *no-information rate* (48.4%). This means that knn achieves the same performance as a naïve classifier that only predicts the majority class.

In summary, the algorithms yield less than 3% higher performance than the no-information rate in classification. This is a relatively poor result given the presence of moderate correlations with the target variable.

### 3.2.1 Feature Importance

Figure 4 shows the feature importance for the best model in the benchmarking, random forests. The basic emotions *joy, disgust,* and *trust,* as well as *negative* and *positive* emotional valence yield the highest importance. In contrast, negative emotions *anger*, *fear* and *sadness* show very low scores corresponding to a negligible feature importance.

### 3.3 Study 2

Study 2 applies tools from XAI methodology to uncover the prediction mechanism of Study 1's best model, random forests. The goal is to understand the prediction mechanism by looking at the instance level and per feature.

The following analyses were computed by averaging over five unseen observations for 1 vs. 5 stars given in the reviews rating.

#### 3.3.1 Comparison of Additive Variable Attributions

The contribution of the features to the model's prediction is shown in Figure 5. Overall on feature influence, some consistency can be noted between the local and global level – the top 3 features of the local attributions analysis for the 1-star rating (joy, disgust, positive) and 5-star rating (joy, negative, disgust) are within the top 4 of global feature importance (Figure 4).

The local attributions analysis shows that the feature joy had the highest impact for the lowest (1 star) and highest (5-star) review rating. A relatively low disgust score led to a high decrease (-0.296) for the 1-star rating. In contrast, for the 5-star rating, the absence of any negative, disgust, sadness, and anger sentiments (0) led to an increase of the response value. These effects are all plausible in direction. For the 1-star rating, however, their effect size is not sufficient to result in a correct prediction but results in a much higher response score (2.8).

A closer look reveals several inconsistencies in the prediction mechanism. First of all, in the 5-star ratings, a positive score of fear (negative emotion) leads to an increase, whereas a positive score for trust (positive emotion) leads to a decrease in the response score. Second, in the 1-star rating, a lower fear score (.0370 vs. .0526) leads to a response decrease (-.082) vs. response increase (+.045), and a lower trust score (.1944 vs. .2105) leads to a stronger response decrease (-.050 vs. -.032). Nevertheless, these effect sizes are relatively small.

In summary, the most important two features (joy and disgust for 1-star rating, joy and negative for 5-star rating) are plausible in direction and effect size. The reason why the prediction is so far off the 1-star rating may lie in the very high intercept (4.085).

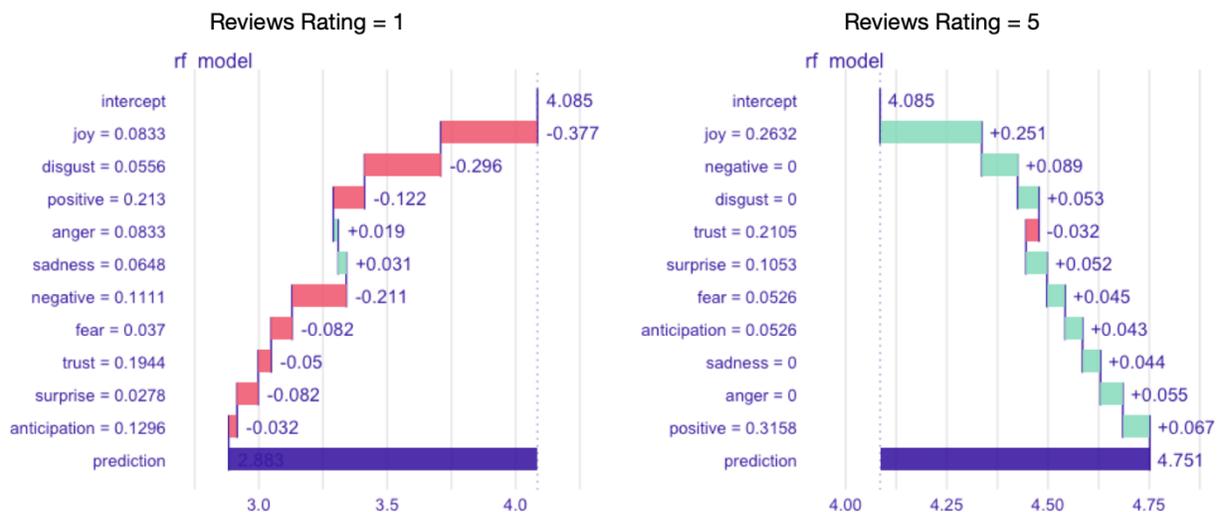

Figure 5: Additive Variable Attributions

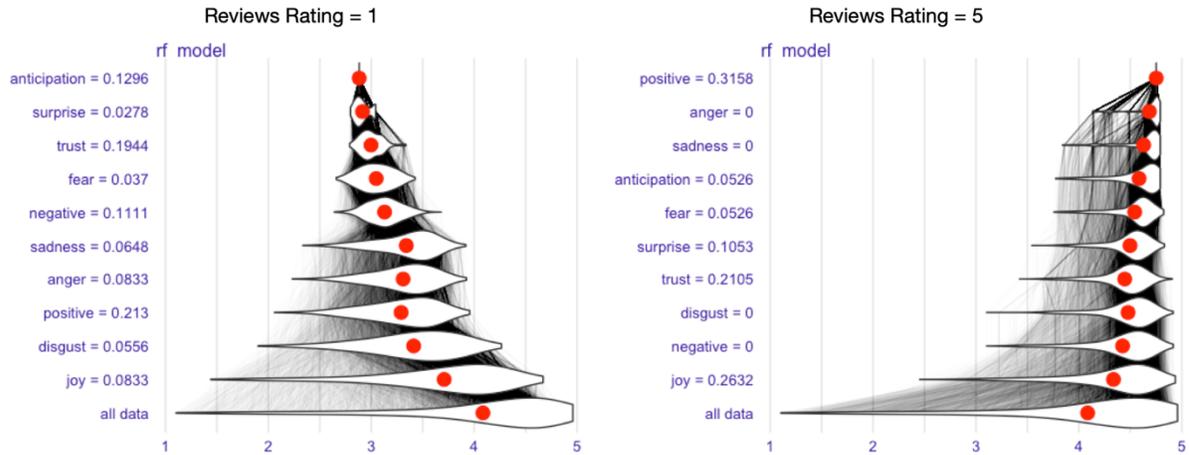

Figure 4: Distribution of all predictions vs. local instance

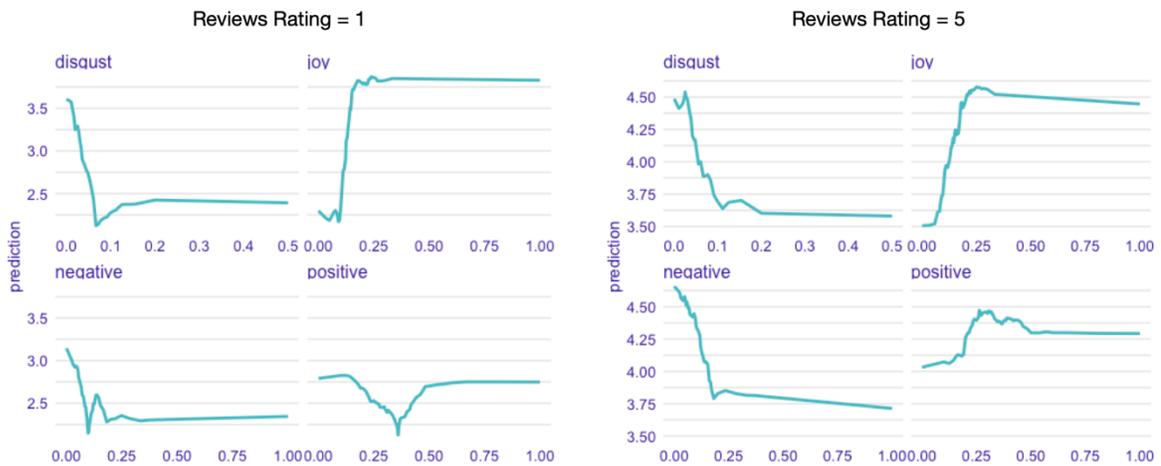

Figure 5: Ceteris paribus plots

Figure 6 shows the distribution of predictions. For the feature scores of the 5-star observations, the prediction distributions quickly converge to higher response scores. However, the mean predictions (red) converge slower, with one inconsistent direction (positive trust value leads to response decrease). For the 1-star observations, the prediction distributions and mean predictions decrease noticeably only with the sentiments joy, disgust, positive, and negative. Overall, the decrease of response value converges to a score (2.8) that lies much higher above the ground truth (1).

*3.3.2 What-if Analysis*
Figure 7 shows the ceteris paribus plots representing a what-if analysis holding all other variables constant. For both the 1-star and the 5-star rating observations, the relationships are correctly reflected for disgust (negative slope) and joy (positive slope). However, the relationship for the sentiment positive (positive slope) is incorrectly reflected for the 1-star review rating (V-shape). The latter also shows a very modest negative slope for the sentiment negative.

Overall, for all features, the sensitivity is very high in a small range near zero, and then remains constant. This aspect demonstrates the limitation of the features' predictive value.

## 4. DISCUSSION
This work aimed to shed light into the prediction mechanism of machine learning models. Several XAI methods uncovered the importance, distribution, direction and effect size of all features for particular instances. These details showed the correct and incorrect functioning of the learned prediction mechanism.

In conclusion, two suggestions to improve the prediction can be made: First, several features of low global and local importance can be discarded (anticipation, fear, sadness, surprise). Second, a balanced dataset would enhance the prediction of the low ratings.

## 5. ACKNOWLEDGMENTS
This research was supported by the Yonsei University Faculty Research Fund of 2019-22-0199.

## 6. REFERENCES

[1] D. Tsui, "Predicting Stock Price Movement Using Social Media Analysis," 2017.

[2] A. Mudinas, D. Zhang, and M. Levene, "Market Trend Prediction using Sentiment Analysis: Lessons Learned and Paths Forward," *Arxiv*, 2019.

[3] Y. Kim, M. Kang, and S. R. Jeong, "Text mining and sentiment analysis for predicting box office success," *KSII Trans. Internet Inf. Syst.*, vol. 12, no. 8, pp. 4090–4102, 2018, doi: 10.3837/tiis.2018.08.030.

[4] S. De Kok, L. Punt, R. Van Den Puttelaar, K. Ranta, K. Schouten, and F. Frasincar, "Review-level aspect-based sentiment analysis using an ontology," *Proc. ACM Symp. Appl. Comput.*, pp. 315–322, 2018, doi: 10.1145/3167132.3167163.

[5] Y. Yu, "Aspect-based Sentiment Analysis on Hotel Reviews," *Arxiv Prepr.*, p. 10, 2016.

[6] P. Ekman, "Are there basic emotions?," *1Psychological Rev.*, vol. 99, no. 3, pp. 550–553, 1992, doi: 10.4081/jear.2011.169.

[7] H. Saarimäki *et al.*, "Discrete Neural Signatures of Basic Emotions," *Cereb. Cortex*, vol. 26, no. 6, pp. 2563–2573, 2016, doi: 10.1093/cercor/bhv086.

[8] E. Diener and R. A. Emmons, "The independence of positive and negative affect," *Clin. Endocrinol. (Oxf).*, vol. 41, no. 4, pp. 421–424, 1994, doi: 10.1111/j.1365-2265.1994.tb02571.x.

[9] D. V. Carvalho, E. M. Pereira, and J. S. Cardoso, "Machine learning interpretability: A survey on methods and metrics," *Electron.*, vol. 8, no. 8, pp. 1–34, 2019, doi: 10.3390/electronics8080832.

[10] C. Rudin, "Stop explaining black box machine learning models for high stakes decisions and use interpretable models instead," *Nat. Mach. Intell.*, vol. 1, no. 5, pp. 206–215, Nov. 2019, doi: 10.1038/s42256-019-0048-x.

[11] S. M. Mohammad and P. D. Turney, "Crowdsourcing a word-emotion association lexicon," *Comput. Intell.*, vol. 29, no. 3, pp. 436–465, 2013, doi: 10.1111/j.1467-8640.2012.00460.x.

[12] S. M. Mohammad and P. D. Turney, "Emotions evoked by common words and phrases: using mechanical turk to create an emotion lexicon," *CAAGET '10 Proc. NAACL HLT 2010 Work. Comput. Approaches to Anal. Gener. Emot. Text*, no. June, pp. 26–34, 2010.

[13] M. Kuhn, "caret: Classification and Regression Training." 2018.

[14] A. Karatzoglou, A. Smola, K. Hornik, and A. Zeileis, "kernlab -- An S4 Package for Kernel Methods in R," *J. Stat. Softw.*, vol. 11, no. 9, pp. 1–20, 2004.

[15] T. Therneau and B. Atkinson, "rpart: Recursive Partitioning and Regression Trees." 2019.

[16] A. Liaw and M. Wiener, "Classification and Regression by randomForest," *R News*, vol. 2, no. 3, pp. 18–22, 2002.

[17] B. Greenwell, B. Boehmke, J. Cunningham, and G. B. M. Developers, "gbm: Generalized Boosted Regression Models." 2019.

[18] F. Doshi-Velez and B. Kim, "Towards A Rigorous Science of Interpretable Machine Learning," no. Ml, pp. 1–13, 2017.

[19] P. Biecek, "DALEX: Explainers for Complex Predictive Models in R," *J. Mach. Learn. Res.*, vol. 19, no. 84, pp. 1–5, 2018.

[20] A. Fisher, C. Rudin, and F. Dominici, "Model Class Reliance: Variable Importance Measures for any Machine Learning Model Class, from the 'Rashomon' Perspective," *J. Mach. Learn. Res.*, vol. 20, 2019.